\documentstyle[12pt]{article}
\author{P. M. Nadolsky, S. M. Troshin, N. E. Tyurin\\
Institute for High Energy Physics\\
Protvino, Moscow Region, 142284 Russia}
\title{ Preasymptotic Nature of Hadron
Scattering\\ vs Small-$x$ HERA Data \rm}
\date{}
\begin{document}
\maketitle
\begin{abstract} We emphasize that recently observed
regularities  in hadron interactions and deep-inelastic scattering
are of preasymptotic nature and it is impossible to make conclusions
on the true asymptotic behavior of observables without unitarization
procedure.  Unitarization is important and changes  scattering
picture drastically.
\end{abstract}

The existence of the regular calculational method for hard
processes inspires one to promptly apply  the parton model also to
calculation of cross sections for soft processes. The problems of
this extension  are notorious.  They are related both to the increase of the
effective coupling constant and the phenomenon of the spontaneous breaking of
chiral symmetry.  The chiral symmetry breaking results, in
particular, in generation of large quark masses and in appearance of quark
condensates.  Nevertheless, the
leading logarithm approximation in the framework of perturbative QCD
was used to derive several results for soft processes, in
particular, the value of hard Pomeron intercept
$1+\Delta\simeq 1.5$~\cite{lipa}. This value seems to find confirmation in
the deep--inelastic scattering data at small $x$, obtained recently at
HERA~\cite{hera}.

 Having in mind the new data from HERA and much discussion around hard
Pomeron \cite{hera}, in this note we would like to emphasize  that the
scattering of hadrons at energies lower than $\sqrt{s}\sim 0.5$ TeV
is of preasymptotic nature. Thus, at this range of energies
we have not any restrictions on the possible behavior of cross--sections,
following from asymptotic theorems.

In general, only an
approach, explicitly taking into account the unitarity, allows
one to accurately distinguish between
the asymptotic and preasymptotic regimes of scattering.
In Refs. \cite{yus,nuov} we have used the
notions of effective chiral quark models for the description of elastic
scattering at small and large angles.  This description is
based on the results of the effective Lagrangian approach and
accounts various aspects of hadron dynamics. For example, massive quarks
appear as quasiparticles, i.e. as current quarks surrounded by the clouds of
quark--antiquark pairs of different flavors.
Besides  mass, quark acquires non-trivial internal structure  and finite
size.  Quark radii are determined by the radii of the clouds.  Strong
interaction radius of the massive quark  $Q$ is determined by its Compton
wavelength:
\begin{equation} r_Q=\xi /m_Q, \label{rq}
\end{equation}
where the
constant $\xi$ is universal for different flavors.
The quark formfactor $F_Q(q)$
is taken in the dipole form, viz
\begin{equation} F_Q(q)\simeq
(1+\xi^2{\vec{q}}^{\,2}/m_Q^2)^{-2}, \label{ff}
\end{equation}
and the corresponding quark matter distribution $d_Q(b)$ is
of the form \cite{nuov}
 \begin{equation} d_Q(b)\propto \exp(-{m_Qb}/{\xi}).  \label{bf}
\end{equation}
Quantum numbers of the constituent quarks are the same as the
quantum numbers of current quarks due to conservation of the corresponding
currents in QCD (we do not concern here the axial-vector currents related to
spin degrees of freedom).

A common feature of chiral models \cite{ball} is the
representation of a baryon as an inner core, carrying the baryonic
charge, and an outer condensate, surrounding this core \cite{isl}.
Following these observations, it is natural to represent a hadron as
consisting  of  the  inner  region, where constituent quarks are located,
and the outer region, filled  with the quark condensate~\cite{nuov}.  This
picture implies  that  the first stage of the hadron collision is
determined by the overlap
and interaction of peripheral condensates. In the overlapping region
the condensates interact and, as a
result, virtual massive quarks appear.
In other words, nonlinear field couplings  transform
kinetic  energy of condensates into the internal energy  of dressed quarks
(see the arguments
for this mechanism in \cite{carr} and references therein for the earlier
works). Of course, the number of such  quarks fluctuates.  The average number
 of quarks is proportional  to the convolution  of the condensate
distributions $D^H_c$ of colliding hadrons:
\begin{equation} \tilde N(s,b)
\simeq {\cal N}(s)\cdot D^A_c \otimes D^B_c, \end{equation}
where the function  ${\cal N}(s)$  is
determined  by thermodynamics  of the kinetic energy transformation.
To estimate
${\cal N}(s)$ one may assume that it has  maximal possible
energy dependence,
\begin{equation} {\cal N}(s) \simeq \kappa  \frac{(1-\langle
x_Q\rangle )\sqrt{s}}{m_Q}, \label{kp} \end{equation}
where $\langle x_Q
\rangle $ is the average fraction of the energy carried by constituent
quarks, $m_Q$ is the mass of the constituent quark.

In the model  the constituent quarks, located in the central part of the
hadron, are supposed to scatter in a quasi-inde\-pen\-dent way both by the
produced virtual massive quarks  and by
other constituent quarks.  The
scattering  amplitude  of the constituent quark, smeared over its
longitudinal
momentum, may be then represented in the form
\begin{equation}
\langle f_Q (s,b) \rangle = [\tilde
N(s,b) + N-1]\langle V_Q(b) \rangle,
\end{equation}
where
$N=N_1+N_2$  is  the total  number  of  constituent quarks  in colliding
hadrons, and $\langle V_Q(b) \rangle$ is  the averaged amplitude  of single
quark-quark scattering.

In this approach the elastic scattering amplitude is constructed
as a solution  of  the equation~ \cite{logu}
\begin{equation} F = U + iUDF,
\label{xx}
\end{equation}
which is presented here in the operator form.
This equation allows  one to satisfy unitarity, provided the
 inequality
\begin{equation}
\mbox{Im\ } U(s,b) \geq 0
\end{equation}
is fulfilled. The function $U(s,b)$  (generalized  reaction
matrix)~\cite{logu} --- the basic dynamical quantity of
this approach --- is
chosen as a product of the averaged quark amplitudes
\begin{equation}
U(s,b) = \prod^{N}_{Q=1} \langle f_Q(s,b)\rangle,
\end{equation}
in accordance  with the assumed quasi-independent  nature  of  valence
quark scattering.

The $b$--dependence of the function $\langle f_Q \rangle$, related to
 the quark formfactor $F_Q(q)$, has a simple form $\langle
f_Q\rangle\propto\exp(-m_Qb/\xi )$.

Following the above considerations, the generalized
reaction matrix in a pure imaginary case is represented in the form
\begin{equation} U(s,b) = iG(N-1)^N \left [1+\alpha
\frac{\sqrt{s}}{m_Q}\right]^N \exp(-Mb/\xi ), \label{x}
\end{equation}
where $M =\sum^N_{q=1}m_Q$, $G=\prod_{Q=1}^N g_Q$ and
$\alpha=\kappa (1-\langle x_Q\rangle)/(N-1)$. This expression allows
one to get the scattering amplitude  as  a  solution  of Eq.
\ref{xx}, reproducing the main regularities observed in  elastic
scattering at small and large angles.

In the impact parameter representation the scattering amplitude may be
written in the form:
\begin{equation} F(s,b)=U(s,b)[1-iU(s,b)]^{-1}.
\label{12}
\end{equation}
This is the solution of Eq.\ref{xx} at $s\gg
4m^2$. Note that the more familiar way to provide the direct channel
unitarity consists in the representation of the scattering amplitude in
 the eikonal form
\[ F(s,b)=\frac{i}{2}\left(1-e^{i\chi
(s,b)}\right), \]
where $\chi(s,b)$ is the eikonal function, related
to the function $U(s,b)$ by the equation:
\[
\chi(s,b)=i\ln\frac{1-iU(s,b)}{1+iU(s,b)}.  \]

At moderate energies $s\ll s_0$, where $\sqrt{s_0}=m_Q/\alpha$ (note
that the magnitude of $\alpha$ can be derived from the numerical analysis
\cite{yus} and is about $1.5\cdot 10^{-4}$, hence  $\sqrt{s_0}\simeq
2$ TeV), the function $U(s,b)$ can be represented in the form
\begin{equation} U(s,b) = i\tilde{g} \left [1+N\alpha
\frac{\sqrt{s}}{m_Q}\right] \exp(-Mb/\xi ),
\label{xl}
\end{equation}
where $\tilde{g}=G(N-1)^N$.  At very high energies $s\gg s_0$ we
can neglect the energy independent term in Eq.\ref{x} and rewrite
the expression for $U(s,b)$ as  \begin{equation}
U(s,b)=i\tilde{g}\left(s/m^2_Q\right)^{N/2}\exp (-Mb/\xi ).
\label{xh} \end{equation}

 Calculation of the scattering amplitude is based on the impact
parameter representation \[
F(s,t)=\frac{s}{2\pi^2}\int\limits^\infty_0d\beta
F(s,\beta)J_0(\sqrt{-\beta t}),\, \beta=b^2 \] and the analysis of
singularities of $F(s,\beta )$ in the complex $\beta $--plane.

Besides the energy dependence of these observables, we will emphasize
its dependence on geometrical characteristics of non--perturbative
quark interactions.

The total cross--section has the following energy and quark mass
dependencies:
\begin{equation} \sigma _{tot}(s)=\frac{\pi  \xi
^2}{\langle m_Q \rangle ^2}\Phi (s,N), \label{y}
\end{equation}
where
$\langle m_Q \rangle=\frac{1}{N}\sum_{Q=1}^N m_Q $ is the mean value
of the constituent quark masses in the colliding hadrons.  The
function $\Phi$ has the following behavior:  \begin{equation} \Phi
(s,N)=\left\{ \begin{array}{cl} \left(8\tilde{g}/N^2\right) \left
[1+N\alpha\sqrt{s}/m_Q\right], & s\ll s_0,\\[2ex] \ln ^2 s, & s\gg
s_0.  \end{array} \right. \label{yy} \end{equation} Thus, at
asymptotically high energies the model provides \[ \lim_{s\rightarrow
\infty}\frac{\sigma_{tot}(\bar a b)} {\sigma_{tot}(ab)}=1.  \]

The preasymptotic  rise of the total
cross--sections, linear with $\sqrt{s}$, is in agreement with the
experimental data say up to $\sqrt{s}\sim 0.5 $ TeV.  In Fig.1--3
the dependence
\begin{equation} \sigma_{tot}=A+B\sqrt{s}
\label{stot}
\end{equation} is compared with the experimental data for $\bar p p$, $pp$,
$K^\pm p$ and $\pi^\pm p$ interactions. It is interesting to note
that these simple fits with two free parameters $A$ and $B$  indicate
a possible intersection of particle and antiparticle total
cross--sections, i.e. simulate the Odderon effect.

The inelastic cross-section can be calculated in the model
explicitely, viz
\begin{equation} \sigma _{inel}(s)=\frac{8\pi \xi
^2}{N^2 \langle m_Q \rangle ^2} \ln \left[1+\tilde{g}(1+\frac{\alpha
\sqrt{s}}{m_Q})^N\right].
\label{s}
\end{equation}
At asymptotically
high energies the inelastic cross--section growth is as follows:
\begin{equation} \sigma _{inel}(s)=\frac{4\pi \xi ^2}{N \langle m_Q
\rangle ^2} \ln s. \label{ss} \end{equation}

At $s\gg s_0 $ the dependence of the hadron interaction radius $R(s)$
and the ratio $\sigma_{el}/\sigma_{tot}$ on $s$
is given by the following equations:
 \begin{eqnarray}
 R(s) & = & \frac{\xi }{2 \langle m_Q \rangle} \ln s, \label{rr}\\
\frac{\sigma_{el}(s)}{\sigma _{tot}(s)} & = & 1-\frac{4}{N \ln s}.
 \label{rs} \end{eqnarray}
It is important to note here that such
behavior of the ratio $\sigma_{el}/\sigma_{tot}$ and $\sigma
_{inel}(s)$ is a result of the self--damping of inelastic channels
\cite{bla} at small impact distances. Numerical estimates \cite{yus}
show that the ratio $\sigma_{el}(s)/\sigma
_{tot}(s)$ reaches the asymptotic value 1 at extremely
high energy $\sqrt{s}=500$ TeV.

 Slower relative increase of the inelastic cross-section at high energies
 is due to the  fact  that the inelastic overlap function $\eta  (s,b)$
becomes peripheral, and  the whole  picture
 corresponds to the antishadow scattering at $b < R(s)$ and to the
shadow scattering at $b>R(s)$.

  Such behavior of the contribution of  inelastic channels arises
because  the scattering  amplitude $f(s,b)$ goes
beyond the black disc limit with the growth of energy:
\[
|f(s,b)| > 1/2
\]
at $b< R(s)$ \cite{bld}.
The appearance  of  the region where the scattering process has
antishadow nature is due  to the self-damping of inelastic channels.
Indeed, in the pure  imaginary case $U(s,b)$ arises as a shadow of
inelastic  processes.  However, the increase  of  the  function  $U(s,b)$
due  to  the increase  of   the contributions of inelastic channels
leads  to  the  decrease  of inelastic overlap function  at  $b<R(s)$
and  feedback the elastic channels. This behavior of the
cross-sections is in accord with the lower bound for elastic
cross-section \cite{macd}:
\begin{equation}
\sigma_{el}(s)>\left[\frac{\sigma_{tot}(s)}{36\pi g(s)}\right]
\sigma_{tot}(s), \end{equation}
\[ g(s)=\frac{d}{dt}(\ln\mbox{Im\,}F(s,t))|_{t=0}.  \]

 The  quantitative  analysis  of the experimental data \cite{yus}
 indicates that antishadow  scattering mode starts to develop at
$\sqrt{s}=2$ TeV.  This  result  is  in agreement with the
experimental indications from the CDF data that the elastic amplitude
$f(s,b)$ at $b=0$  already reaches  the  black disc limit
\cite{belf}. The development of the antishadow mode in head-on  $\bar
p p$--collisions at Tevatron could be associated with new phenomena
in  the  central  hadronic  collisions  where  the temperatures are
high and the energy density can be up to  several $GeV/fm^3$.

Thus, unitarization drastically changes the scattering picture:  at
lower energies inelastic channels provide dominant contribution and
scattering amplitude has a shadow origin, while at high energies
elastic scattering dominates over inelasic contribution and the
scattering picture corresponds to the antishadow mode.  The
functional $s$--dependencies of observables also differ
significantly.  For example, $s$--dependence of the total cross-section
at $s\ll s_0$ is described by a simple linear function of $\sqrt{s}$.
It has been shown that such dependence does not contradict the
experimental data for hadron total cross--sections  at least up to
$\sqrt{s}\sim 0.5$ TeV  (see Figs. 1-3 and \cite{yus} for
earlier results).  Such dependence  corresponds to that of the hard
Pomeron with $\Delta=0.5$, however, it is obtained in a different
 approach \cite{nuov}.  This is the preasymptotic dependence and it has
nothing to do  with the true asymptotics of the total cross-sections.
In the model such behavior of hadronic cross--sections reflects
the energy dependence of the number of virtual quarks, generated in the
intermediate transient stage of hadronic interaction.

 The experimental data on $\sigma_{tot}(\gamma p)$  can  also be
 described by Eq.\ref{stot}.  The comparison with the  data is given
 in Fig. 4.  New data from HERA at small $x$ correspond to the c.m.
 energy range of $\gamma^* p$ system $W=50-300$ GeV which, judging
 by the hadronic cross--sections behavior,
should be considered as the preasymptotic energy region.
In this region the rise of hadronic cross-sections is consistent with the
linear dependence on $\sqrt{s}$.

The observed behavior of the structure
function $F_2$ at small $x$ or, equivalently, the dependence of
$\sigma_{tot}(\gamma^*p)$ on $W$ \cite{hera} favors
Eq.~\ref{stot} with $Q^2$-dependent parameters $A$ and $B$ (cf. Fig.  19
in~\cite{hera}, ZEUS data).
In the framework of the model \cite{nuov} we can speculate that this
$Q^2$--dependence reflects the fact that the efficiency of the
transformation of hadron kinetic energy
into the masses of virtual quarks (parameter $\kappa$ in
 Eq.~\ref{kp}) depends on $Q^2$,  $\kappa\rightarrow\kappa
 (Q^2)$, and the function $\kappa(Q^2)$ has the critical behavior with
$Q^2$, viz it has a steep increase at $Q^2\sim 1$ GeV$^2$.
Thus, it seems premature to claim that the
hadronic data and data obtained in deep-inelastic scattering require
two Pomerons -- soft and hard ones \cite{land}.

\subsubsection*{Acknowledgement}
We are grateful to V.A. Petrov for
interesting discussions.

\newpage
\normalsize
\subsubsection*{Figure captions}
\bf Fig. 1  \rm  Total cross--sections of $\bar p p$-- and
$pp$--interactions.  Data from Tevatron at $\sqrt{s}=1.8$ TeV  beyond
 the preasymptotic region have not been included.\\[2ex]
\bf Fig. 2
\rm  Total cross--sections of $\pi^\pm p$--interactions.\\[2ex]
\bf Fig. 3  \rm  Total cross--sections of $K^\pm p$--interactions.\\[2ex]
\bf Fig. 4  \rm  Total cross--section of $\gamma p$--interactions.

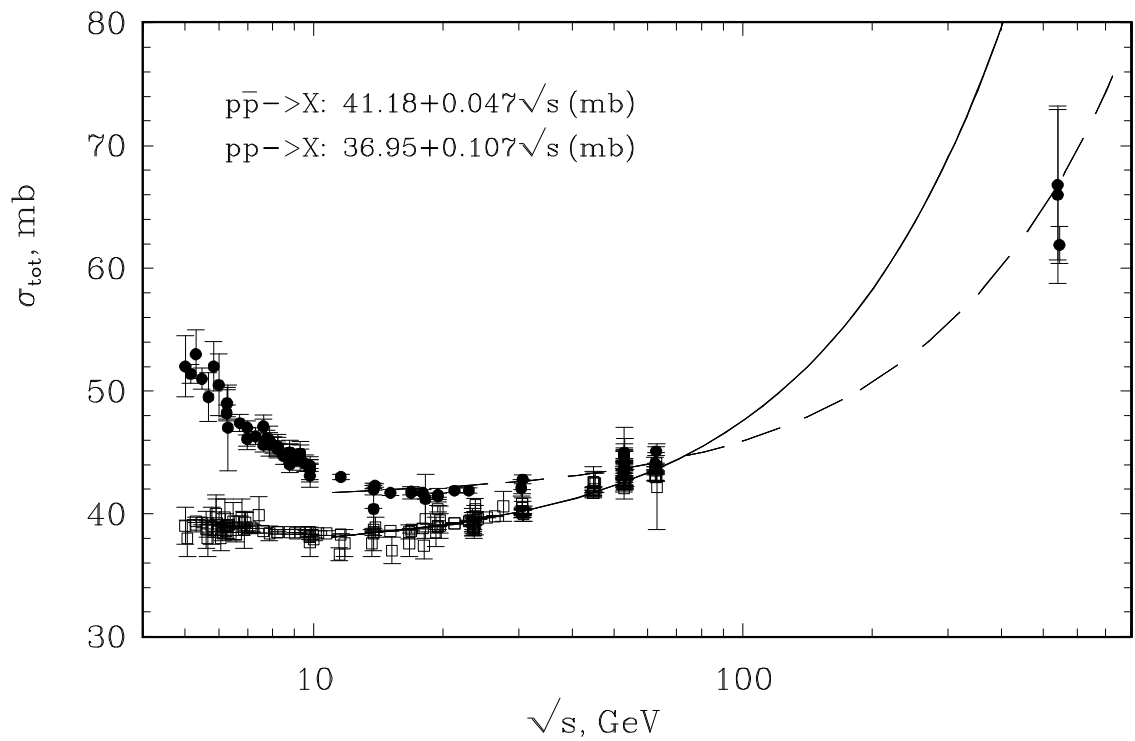
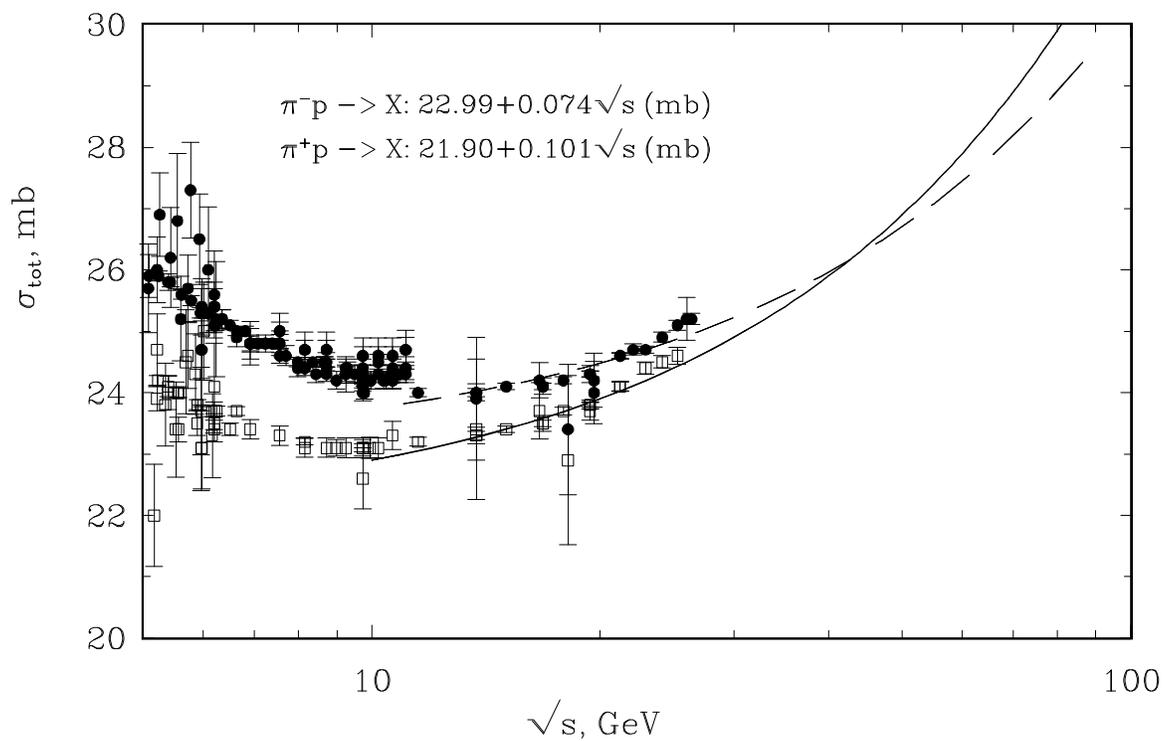
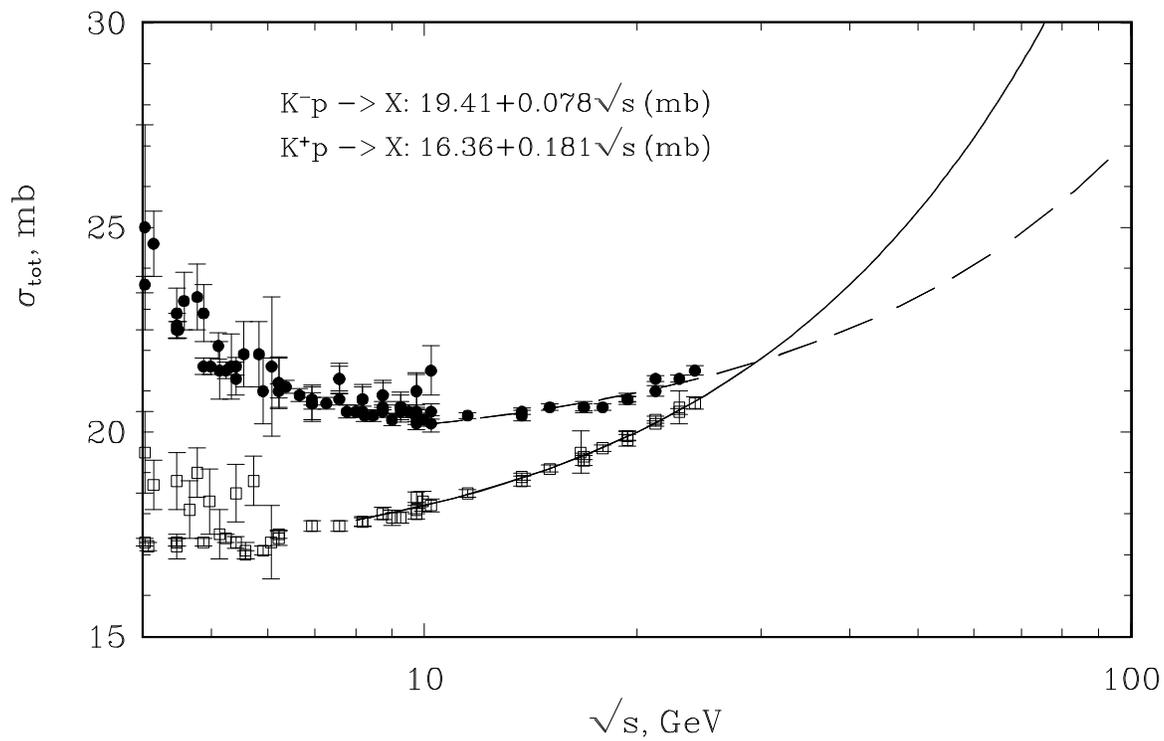
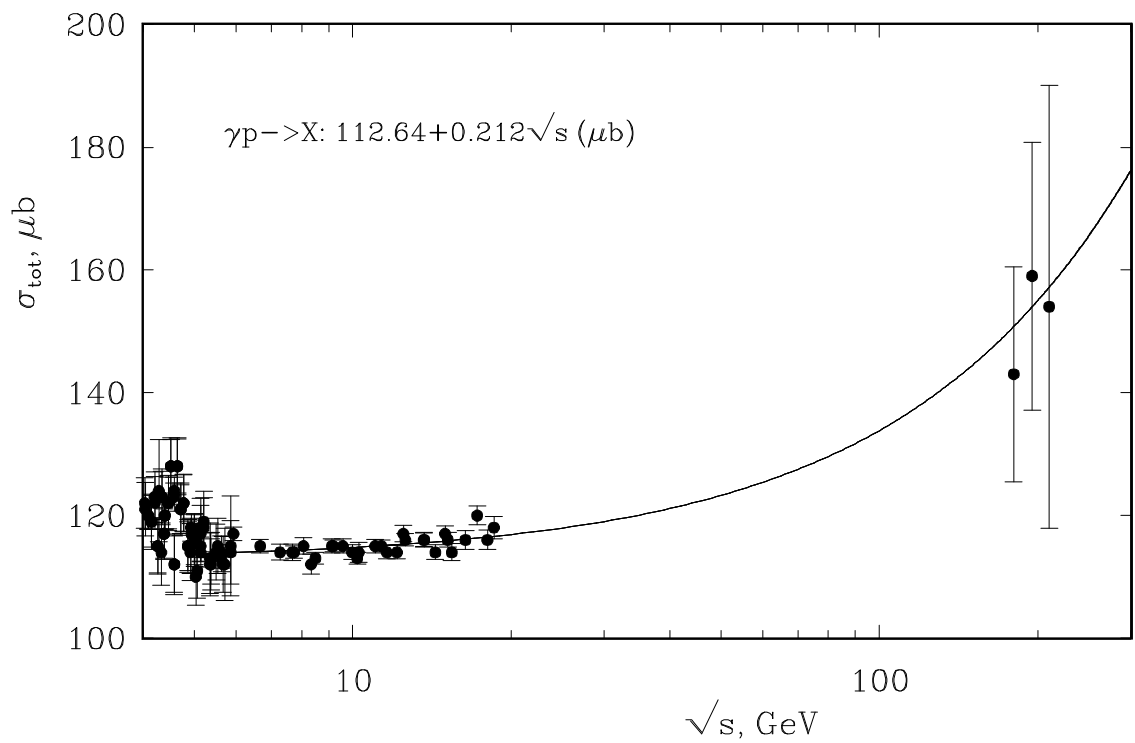
\end{document}